\documentclass{article}

\usepackage{PRIMEarxiv}
\usepackage[utf8]{inputenc}
\usepackage[T1]{fontenc}
\usepackage{graphicx}
\usepackage{booktabs}
\usepackage{array}
\usepackage{longtable}
\usepackage{hyperref}
\usepackage{cite}
\usepackage{amsmath, amssymb}
\usepackage{titlesec}
\usepackage{xcolor}
\usepackage{url}
\usepackage{enumitem}
\usepackage{multirow}
\title{Rethinking Software Engineering for Agentic AI Systems}

\author{
  Mamdouh Alenezi \\
  Saudi Data and Artificial Intelligence Authority (SDAIA) \\
  Riyadh, Saudi Arabia\\
}

\begin{document}
\maketitle

\begin{abstract}
\noindent
The rapid proliferation of large language models (LLMs) and agentic AI systems has generated an unprecedented abundance of automatically synthesized code, fundamentally challenging the traditional software engineering paradigm centered on manual code authorship. This paper investigates the following research question: \textit{Given the increasing volume and quality of code generated by LLMs and agentic AI systems, should software engineering be redefined around orchestration, verification, and human-AI collaboration---and what concrete changes in education, tools, processes, and professional practice would such a shift require?} Through a structured multivocal literature review spanning 23 recent peer-reviewed studies, preprints, and expert surveys, we synthesize evidence across four thematic pillars: the evolving role of the engineer in agentic workflows, verification as the emerging quality bottleneck, empirical assessments of productivity and maintainability, and implications for education and professional practice. Our analysis reveals that code is transitioning from a scarce, laboriously crafted artifact to an abundant, disposable commodity, and that the discipline must consequently reorganize around three core competencies: strategic orchestration of multi-agent workflows, rigorous verification of AI-generated outputs, and structured human-AI collaboration. We present a conceptual framework detailing the concrete transformations required across engineering curricula, development tooling, lifecycle processes, and professional governance. Our findings indicate that this paradigm shift does not replace human engineers but elevates their role toward system-level design, semantic verification, and accountable oversight. The paper concludes with a research roadmap addressing open challenges in verification-first lifecycle adoption, prompt provenance standards, and longitudinal workforce studies.
\end{abstract}


\section{Introduction}
\label{sec:introduction}

For decades, the discipline of software engineering has been anchored in the manual construction of code, where developer productivity is measured by lines written, architectural diagrams drafted, and iterative debugging cycles completed \cite{Sadowski2019, Gul2025}. Established process models---from waterfall to agile---assume that code authorship is the central, rate-limiting activity of the development lifecycle. However, the rapid maturation of large language models (LLMs) and their integration into agentic AI systems has fundamentally disrupted this assumption. Modern LLMs produce functional, syntactically correct code at scale across diverse development tasks, from unit-level function generation to cross-file refactoring and deployment scripting \cite{liang2026dronetest, fan2023llmse, zhong2025llmse}. As a consequence, code is no longer a scarce artifact painstakingly crafted by human hands; it is becoming an abundant, disposable commodity that can be synthesized on demand.

This shift raises a question that strikes at the identity of the discipline. If the primary output of software engineering---source code---can be produced cheaply and at scale by AI systems, what is the enduring value proposition of the software engineer? And how must the discipline's educational foundations, tooling ecosystems, process models, and professional structures adapt to remain relevant and rigorous?

This paper investigates the following research question:

\begin{quote}
\textit{RQ: Given the increasing volume and quality of code generated by LLMs and agentic AI systems, should software engineering be redefined around orchestration, verification, and human-AI collaboration rather than traditional code authorship---and what concrete changes in education, tools, processes, and professional practice would such a shift require?}
\end{quote}

\noindent We decompose this overarching question into four sub-questions to guide our analysis:

\begin{quote}
\textit{RQ1:} How is the role of the software engineer evolving in the context of agentic AI workflows?\\[0.3em]
\textit{RQ2:} What verification challenges arise from the abundance of AI-generated code, and what hybrid approaches address them?\\[0.3em]
\textit{RQ3:} What does the empirical evidence reveal about the productivity, quality, and maintainability effects of AI-augmented development?\\[0.3em]
\textit{RQ4:} What transformations in education, tooling, processes, and professional practice are required to operationalize this paradigm shift?
\end{quote}

To address these questions, we conduct a structured multivocal literature review of 23 recent studies---including peer-reviewed journal articles, conference proceedings, and preprints---selected for their direct relevance to the intersection of LLMs, agentic AI, and software engineering practice. Our analysis synthesizes findings across four thematic pillars: role evolution, verification, empirical evidence, and disciplinary transformation. On the basis of this synthesis, we propose a conceptual framework that maps the core competencies, tooling requirements, process adaptations, and governance structures needed to operationalize an orchestration- and verification-centered model of software engineering.

The contributions of this paper are threefold. First, we provide a thematically organized synthesis of the rapidly growing literature on AI-augmented software engineering, identifying areas of convergence and remaining gaps. Second, we articulate a framework of four core competencies---intent articulation, systematic verification, multi-agent orchestration, and accountable human judgment---that redefine the professional identity of the software engineer. Third, we offer a concrete, multi-level transformation roadmap spanning education, tools, processes, and governance, grounded in empirical evidence and designed to guide both researchers and practitioners.

The remainder of this paper is organized as follows. Section~\ref{sec:methodology} describes our review methodology. Section~\ref{sec:related_work} synthesizes the background and related work across the four thematic pillars. Section~\ref{sec:competencies} presents the proposed framework of new core competencies. Section~\ref{sec:transformations} details the concrete disciplinary transformations required. Section~\ref{sec:discussion} discusses implications, limitations, and threats to validity. Section~\ref{sec:conclusion} concludes with a summary and research roadmap.

\section{Research Approach}
\label{sec:methodology}

This study adopts a structured and systematic approach to synthesizing existing knowledge relevant to the research questions. Given the rapidly evolving nature of AI-augmented software engineering, the analysis draws on a diverse body of sources spanning academic literature, emerging research outputs, and practitioner-oriented insights.

Relevant works were identified through iterative search and screening processes, with selection guided by relevance, conceptual clarity, and contribution to the topic. The collected materials were then analyzed using a qualitative synthesis approach to identify recurring themes, patterns, and areas of convergence.

To ensure alignment with research objectives, the analysis was organized around four key dimensions: the evolving role of the software engineer, verification challenges and approaches, observed impacts on productivity and quality, and implications for education and professional practice. Insights across these dimensions were iteratively refined and consolidated into a coherent thematic structure, which underpins the discussion of related work and informs the conceptual framework and transformation roadmap presented in subsequent sections. The resulting thematic structure organizes Section~\ref{sec:related_work} and informs the framework presented in Sections~\ref{sec:competencies} and~\ref{sec:transformations}.

\section{Background and Related Work}
\label{sec:related_work}

\begin{table}[htbp]
\centering
\caption{The Changing Identity of the Software Engineer}
\label{tab:changing-identity}
\renewcommand{\arraystretch}{1.5}
\begin{tabular}{|l|p{5cm}|p{5cm}|}
\hline
 & \multicolumn{1}{c|}{\textbf{The Traditional Era}} & \multicolumn{1}{c|}{\textbf{The Agentic AI Era}} \\
\hline
Primary Artifact & Laborious Scarcity & Disposable Abundance \\
\hline
Engineer's Core Role & Manual Authorship & Strategic Orchestration \\
\hline
System Bottleneck & Writing Syntax & Verifying Semantics \\
\hline
Definition of Success & Lines of Code / PR Throughput & Decision Velocity \& System Reliability \\
\hline
Pedagogical Focus & Language Memorization & Architectural Trade-offs \\
\hline
\end{tabular}
\end{table}

The question of whether software engineering should be redefined around orchestration, verification, and human-AI collaboration has generated substantial and growing scholarly attention. As summarized in Table~\ref{tab:changing-identity}, recent literature converges on a qualified affirmation: the discipline's center of gravity is shifting toward intent specification, hybrid verification pipelines, and accountable human oversight, though not toward the wholesale abandonment of authorship. This section synthesizes evidence across the four thematic pillars identified in our methodology: role evolution and orchestration (Section~\ref{subsec:orchestration}), verification as the emerging bottleneck (Section~\ref{subsec:verification}), empirical assessments of productivity and quality (Section~\ref{subsec:empirical}), and implications for education and professional practice (Section~\ref{subsec:education}).

\subsection{From Code Authorship to Orchestration and Intent Specification}
\label{subsec:orchestration}

A growing body of work argues that LLMs and agentic AI systems are transforming the fundamental unit of software engineering work from syntactic code production to intent articulation and workflow coordination. Fan et al.~\cite{fan2023llmse} provide a foundational survey of LLMs for software engineering, concluding that while emergent capabilities enable significant creativity across coding, design, and repair tasks, they introduce substantial risks---notably hallucinations and specification drift. Their analysis emphasizes the necessity of hybrid techniques that combine traditional software engineering methods with LLM capabilities to reliably filter incorrect solutions. This framing repositions the engineer not as a primary coder but as a curator who integrates AI outputs with established verification methods.

Roychoudhury~\cite{roychoudhury2025agentic} extends this argument to agentic AI workflows, contending that trustworthy automation requires resolving the persistent software engineering challenge of deciphering and clarifying developer intent. In this vision, AI agents handle micro-level implementation decisions autonomously, while human engineers retain responsibility for specification inference, constraint definition, and final accountability. The paper forecasts that future workflows will embed AI-based verification and validation to manage the combinatorial explosion of automatically generated code. Similarly, Zhong~\cite{zhong2025llmse} reviews LLM applications across requirements analysis, debugging, and operations, concluding that sustainable progress depends on robust human-computer collaboration and the establishment of shared standards.

Surveys of LLM-empowered agentic systems further highlight prompt-based, fine-tuning, and agent-based paradigms that enable autonomous task decomposition, tool usage, and iterative refinement across the development lifecycle~\cite{wang2024agentic}. In this emerging model, engineers define goals, constraints, and non-functional requirements in natural or formal language, then orchestrate specialized agents for subtasks such as requirements formalization, test generation, or deployment scripting. This operational shift aligns with recent proposals for self-evolving, agentic teams that handle end-to-end synthesis, freeing human engineers for value alignment and exception handling in ways that neither waterfall nor agile methodologies originally anticipated.

\subsection{Verification as the Emerging Quality Bottleneck}
\label{subsec:verification}

As code generation becomes commoditized, verification is emerging as the critical differentiator for reliable software systems. Dolcetti and Iotti~\cite{dolcetti2025verification} offer a dual-perspective review that structures the field along two complementary axes: the use of LLMs \textit{as} verification tools (e.g., for test generation, invariant suggestion, or bug localization) and the verification \textit{of} code produced by LLMs. Their analysis identifies hybrid methodologies---integrating LLMs with static analyzers, formal methods, and coverage-guided testing---as the most promising path toward trustworthy AI-augmented development.

This finding aligns with Mohan's~\cite{mohan2025llmdv} work in hardware verification, which proposes modular LLM agents that act as code analyzers, coverage interpreters, and assertion suggesters working alongside human engineers. Experimental results in this domain demonstrate gains in both accuracy and interpretability, though the authors caution that industrial adoption requires robust chaining of generation, analysis, and validation steps. Dedicated frameworks for formal verification of AI outputs, such as those employing Ada/SPARK annotations~\cite{li2024verification}, further demonstrate that without systematic verification pipelines, abundant code risks introducing systemic fragility into production systems.

The consensus across these reviews is substantive: standalone LLM outputs cannot be trusted in isolation for production use. Verification must be treated as a first-class concern---embedded in pipelines that combine prompt engineering, symbolic reasoning, dynamic testing, and human review. This structural shift redefines a central aspect of the engineer's role: from writing code to designing and governing the verification infrastructure itself.

\subsection{Empirical Evidence on Productivity, Quality, and Maintainability}
\label{subsec:empirical}

Empirical studies provide nuanced and sometimes cautionary support for the augmentation thesis. Systematic literature reviews of hundreds of LLM-for-Software-Engineering papers reveal that while over half focus on code generation and maintenance tasks---yielding notable productivity gains---performance drops sharply for interdependent, class-level, or architecturally complex artifacts~\cite{chen2023llm4se}. Global expert surveys corroborate these findings, demonstrating that AI augments repetitive coding tasks but elevates engineer effort toward design, verification, and creative problem-solving~\cite{gupta2024survey}. These results underscore that traditional authorship metrics, such as lines of code written, are becoming unreliable indicators of engineering value.

At the individual level, Weisz et al.~\cite{weisz2024enterprise} report on an enterprise deployment of IBM's watsonx Code Assistant, finding net productivity increases but noting that benefits are not uniformly distributed across users. Their mixed-methods analysis reveals that motivations, task fit, and prior expectations significantly moderate outcomes, highlighting the importance of context-aware orchestration strategies rather than one-size-fits-all AI adoption.

Perhaps the most consequential empirical finding concerns downstream code quality. Borg et al.~\cite{borg2025maintainability} conduct a two-phase controlled experiment with 151 professional developers to assess the maintainability of AI-co-developed code. Phase~1 replicates prior findings: AI assistance yielded a 30.7\% median reduction in task completion time. However, Phase~2---in which different developers evolved the resulting code without AI assistance---revealed no significant differences in completion time or code quality compared to a control group. Bayesian analysis suggests that any quality improvements were at most small and highly uncertain. The authors conclude that while speed gains from AI assistance are plausible, the long-term reliability and maintainability of the resulting software depend on the surrounding process infrastructure---particularly verification, review, and testing---rather than on the generative model alone. This finding directly reinforces the argument that verification and human oversight must be structurally embedded in AI-augmented workflows.

\subsection{Implications for Education and Professional Practice}
\label{subsec:education}

The pedagogical implications of this paradigm shift are addressed by Degerli~\cite{degerli2026education}, who proposes a curriculum adaptation framework for the LLM era. The paper argues that core competencies must migrate from syntax production toward critique, validation, and human-AI stewardship, and that traditional plagiarism-centric academic integrity mechanisms are becoming insufficient in environments where AI-generated code is ubiquitous. Instead, assessment should emphasize process transparency, oral defense, and evidence of reasoning over AI-generated outputs. This aligns with broader calls to teach problem framing, architectural thinking, and verification literacy as foundational skills for next-generation software engineers.

Regarding professional practice, Zhong~\cite{zhong2025llmse} and Alenezi and Akour~\cite{alenezi2025ai} highlight emerging concerns around provenance, security, privacy, copyright, and liability. As engineers increasingly define intent, constrain agents, and validate results rather than writing code directly, organizations must establish governance policies that treat these concerns as core engineering responsibilities rather than ancillary compliance tasks. The literature consistently frames accountability as the irreducible human component in AI-augmented development: engineers become the parties who own residual risk and ensure that AI-generated systems align with ethical, legal, and societal expectations.

\subsection{Synthesis and Gap Identification}
\label{subsec:gaps}

Collectively, this body of work supports a redefinition of software engineering as the discipline of designing, constraining, and validating AI-augmented software production systems while preserving accountability for the resulting artifacts. The strongest evidence favors hybrid, verification-first approaches that embed LLMs within traditional quality assurance pipelines. However, several important gaps remain. First, longitudinal studies examining the effects of curriculum interventions and workforce restructuring on engineering outcomes are scarce. Second, standards for prompt versioning, agent capability declaration, and audit-ready development logs remain underdeveloped. Third, empirical research on team dynamics in multi-agent, human-in-the-loop environments is nascent, with most studies examining dyadic human-AI interactions rather than the complex multi-agent topologies emerging in practice. Addressing these gaps will be essential for a responsible, evidence-based transition to the orchestration paradigm.

To ground this synthesis in the specific studies informing our analysis, Table~\ref{tab:papers} categorizes 12 representative peer-reviewed studies by their alignment with the key thematic pillars.

\renewcommand{\arraystretch}{1.25}
\begin{longtable}{p{0.15\textwidth} p{0.36\textwidth} p{0.42\textwidth}}
\caption{Representative Peer-Reviewed Studies Supporting the AI-Augmented Software Engineering Paradigm}
\label{tab:papers} \\
\toprule
\textbf{Pillar} & \textbf{Study} & \textbf{Key Finding} \\
\midrule
\endfirsthead

\caption[]{Representative Peer-Reviewed Studies Supporting the AI-Augmented Software Engineering Paradigm (Continued)} \\
\toprule
\textbf{Pillar} & \textbf{Study} & \textbf{Key Finding} \\
\midrule
\endhead

\bottomrule
\endfoot

\multirow{4}{0.15\textwidth}{\raggedright Orchestration \& Agentic Workflows} 
& Comprehensive Evaluation of LLMs on SE Tasks \cite{chen2023llm4se} 
& Models with identical success rates exhibited 53$\times$ cost differences and vastly different tool-call volumes (917 vs.\ 3), exposing critical inefficiencies in LLM tool use. \\
\addlinespace
& Comparison of LLM-based Agent Configurations \cite{wang2024agentic} 
& No single agent configuration was universally optimal; effectiveness depended on task and domain context, underscoring the need for adaptive orchestration. \\
\addlinespace
& MOSAICO: AI-agent Communities \cite{mosaico2024} 
& Specialized, collaborating AI agents managed by a central platform overcame individual LLM limitations, achieving higher aggregate reliability. \\
\addlinespace
& HAI-Eval: Human-AI Synergy \cite{haieval2024} 
& Neither humans nor AI alone solved complex tasks effectively (0.67--18.89\% success); collaboration more than tripled performance to 31.11\%. \\
\midrule

\multirow{4}{0.15\textwidth}{\raggedright Verification \& Trustworthiness} 
& Dual Perspective Review: LLMs and Code Verification \cite{dolcetti2025verification} 
& Systematic mapping of LLMs as verification tools and as subjects of verification, identifying hybrid approaches as most promising. \\
\addlinespace
& QA of LLM-generated Code \cite{qa2024} 
& Revealed an academia--industry mismatch: academia prioritizes security and performance; industry is more concerned with maintainability and technical debt. \\
\addlinespace
& Vulnerability Detection: Formal to Hybrid \cite{hybrid2024} 
& Hybrid techniques combining formal rigor with LLM scalability enhanced verification coverage beyond either approach alone. \\
\addlinespace
& Rethinking Autonomy: SAFE-AI Framework \cite{safeai2024} 
& Proposed guardrails, sandboxing, and human-in-the-loop systems (SAFE-AI: Safety, Auditability, Feedback, Explainability) for autonomous AI in development. \\
\midrule

\multirow{4}{0.15\textwidth}{\raggedright Education \& Practice} 
& Coding With AI: Practices to Education \cite{edu2024} 
& Found code review to be the new development bottleneck; argued for educational emphasis on problem-solving and architectural thinking. \\
\addlinespace
& Lost in Code Generation \cite{lost2024} 
& Argued that software models will shift from upfront blueprints to post-hoc recovery artifacts mediating between human intent and AI output. \\
\addlinespace
& Code Writers to Code Curators \cite{cefr2024} 
& Introduced a six-stage framework for teaching ``receptive programming''---the ability to read, evaluate, and integrate AI-generated code. \\
\addlinespace
& GenAI and Empirical SE \cite{paradigm2024} 
& Argued that AI as an active collaborator changes the nature of development artifacts, requiring re-evaluation of empirical research methods. \\
\bottomrule
\end{longtable}

\section{The New Core Competencies of Software Engineering}
\label{sec:competencies}

The evidence synthesized in the preceding section reveals that the traditional model of software engineering---centered on manual code construction and process compliance---is giving way to a new professional identity organized around systemic integrity rather than syntactic production. This section articulates four core competencies that, we argue, define the emerging role of the software engineer in AI-augmented development environments.

\subsection{Intent Articulation and Architectural Control}

\begin{figure}[ht!]
  \centering
  \includegraphics[width=0.93\textwidth]{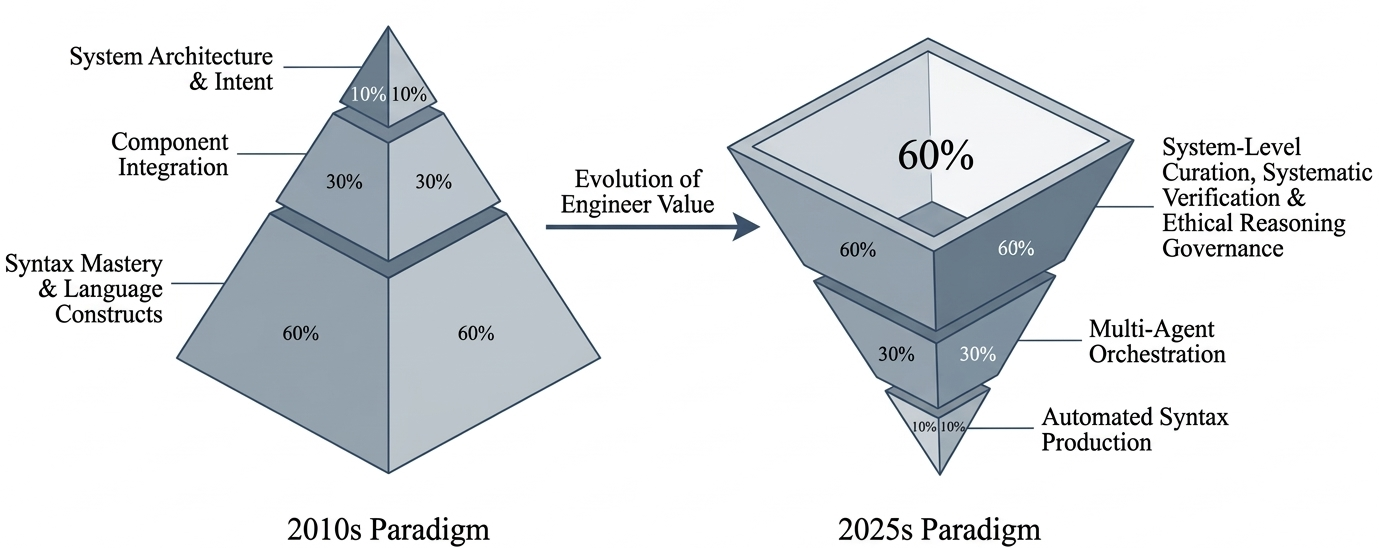}
  \caption{The Inversion of the Engineering Value.}
  \label{fig:engieeringvalue}
\end{figure}

The most fundamental shift in engineering practice is from writing code to precisely specifying \textit{what} should be built and \textit{why} (see Figure~\ref{fig:engieeringvalue}). In an orchestration-centered paradigm, engineers must establish the overall system architecture, define strategic direction, and decompose complex requirements into specifications that AI agents can execute reliably. This competency encompasses requirements formalization, constraint specification in natural or formal languages, and the ability to translate stakeholder needs into machine-actionable directives. The engineer acts not as a line-by-line implementer but as a navigator, steering AI agents toward value-aligned outcomes by articulating boundary conditions, quality attributes, and design trade-offs that generative models cannot infer from training data alone~\cite{roychoudhury2025agentic}.

\subsection{Systematic Verification and Quality Assurance}

As AI-generated code fails differently from human-written code---often syntactically correct but semantically flawed, subtly insecure, or inconsistent with architectural intent---verification has become the new rate-limiting activity in the development lifecycle. Engineers must develop fluency in designing verification pipelines that combine static analysis, dynamic testing, formal methods, and AI-assisted proof generation to catch failure modes specific to generative outputs~\cite{dolcetti2025verification, li2024verification}. This competency extends beyond traditional testing to include semantic validation against specifications, cross-model consistency checking, and the governance of verification infrastructure. The empirical finding that downstream maintainability depends on surrounding process quality rather than on the generative model itself~\cite{borg2025maintainability} underscores that verification is not an optional adjunct but a structural prerequisite for trustworthy AI-augmented development.

\subsection{Multi-Agent Orchestration}

Managing ensembles of specialized AI agents is emerging as a primary engineering task, distinct from both traditional project management and conventional tool use. Orchestration involves coordinating agent workflows, providing contextual grounding, resolving inter-agent conflicts, allocating computational resources, and ensuring that agent outputs converge toward overarching system goals~\cite{wang2024agentic, mosaico2024}. This competency requires fluency in multi-agent communication patterns, prompt engineering, agent capability assessment, and iterative refinement loops. The evidence that no single agent configuration is universally optimal~\cite{wang2024agentic} implies that orchestration itself must be adaptive, with engineers selecting and composing agent teams based on task characteristics, domain constraints, and quality requirements.

\subsection{Human Judgment and Accountability}

The engineer provides the essential human context that AI systems inherently lack: business logic interpretation, user experience sensitivity, ethical reasoning, risk tolerance calibration, and regulatory awareness. As the literature consistently frames accountability as the irreducible human component~\cite{zhong2025llmse, alenezi2025ai}, engineers in the orchestration paradigm become the parties who own residual risk, make final deployment decisions, and ensure alignment with legal and societal expectations. This competency is not merely a soft skill but a formal responsibility: organizations must establish governance structures in which human engineers serve as the accountable decision-makers at critical junctures in AI-augmented pipelines, including architecture sign-off, security validation, and production deployment authorization~\cite{safeai2024}.

\section{Concrete Transformations Across the Discipline}
\label{sec:transformations}

Realizing the competency framework outlined in the preceding section requires coordinated, multi-level transformations across education, tooling, lifecycle processes, and professional governance. These changes must be implemented cohesively to prevent fragmentation and to ensure that the abundance of AI-generated code translates into systemic reliability rather than systemic risk. The transition parallels earlier paradigm shifts---from assembly to high-level languages, from on-premises to cloud-native architectures---in which the economics of production changed and the profession reorganized around higher-value human strengths: vision, judgment, system-level reasoning, and accountability~\cite{russinovich2026redefining, banh2025copiloting}. This section details the concrete transformations required along each of four dimensions.

\subsection{Education: From Syntax Mastery to Systems Curation}
\label{subsec:edu_transform}

Engineering curricula must pivot from language-specific syntax training toward foundational principles of systems thinking, verification literacy, and human-AI collaboration. The evidence from Degerli~\cite{degerli2026education} and related educational studies~\cite{edu2024, cefr2024} indicates that the core pedagogical objective should shift from teaching students to produce code to teaching them to critically evaluate, integrate, and govern AI-generated artifacts. This shift is made urgent by the ``AI drag'' phenomenon identified by Russinovich and Hanselman~\cite{russinovich2026redefining}: while AI coding assistants provide a substantial productivity boost to senior engineers who possess the contextual knowledge to direct agents effectively, they simultaneously impose a productivity drag on early-in-career (EiC) developers who lack the experience to steer, verify, and integrate AI output. If curricula do not adapt, new graduates risk entering the workforce unable to perform the very oversight tasks that define the redefined engineering role.

Concretely, this entails restructuring introductory courses around architecture, domain modeling, design patterns, and trade-off analysis rather than memorization of language constructs, while embedding hands-on training in formal methods and automated testing throughout the curriculum. As Cruz et al.~\cite{cruz2025redefining} document through qualitative interviews with IT professionals, developers are already adapting to AI-augmented environments by engaging in continuous upskilling, prompt engineering, interdisciplinary collaboration, and heightened ethical awareness; curricula must prepare students for these realities from day one. Prompt engineering---understood as the precise articulation of intent, constraint specification, and iterative refinement---should be treated as a foundational development competency rather than a peripheral skill. Indeed, Banh et al.~\cite{banh2025copiloting} identify through a Grounded Theory analysis of practitioner interviews that the most prevalent use case for generative AI in software engineering is the actual coding task, yet the sociotechnical challenges of integration---including trust calibration, contextual fit, and over-reliance---remain substantial barriers that only disciplined training can address.

Project-based coursework should provide early and continuous exposure to realistic AI-assisted workflows, requiring students to manage AI teammates, negotiate shared mental models, and document their reasoning processes. Critically, curricula must also preserve foundational skills through deliberate ``AI-free'' exercises that build the deep understanding necessary for effective AI oversight~\cite{russinovich2026redefining}. Assessment mechanisms must evolve correspondingly: process transparency, oral defense, and evidence of critical reasoning should supplement or replace traditional code-submission-based evaluation to address the insufficiency of plagiarism-centric integrity mechanisms in AI-saturated environments~\cite{degerli2026education}. Finally, responsible AI and ethics---including data privacy, algorithmic bias, accountability frameworks, and societal impact assessment---must be woven throughout the curriculum as a cross-cutting concern rather than isolated in a single elective~\cite{cruz2025redefining}.

The preceptor model proposed by Russinovich and Hanselman~\cite{russinovich2026redefining}, in which senior engineers are paired with EiC developers in real product teams, offers a promising bridge between academic preparation and industrial practice. Universities and industry must collaborate to ensure that today's expertise becomes tomorrow's intuition, preventing the talent-pipeline collapse that results from treating AI as a replacement for, rather than an augmentation of, the learning process.

\begin{table}[htbp]
\centering
\caption{Evolving Engineering Education \& Assessment}
\label{tab:education-assessment}
\small
\renewcommand{\arraystretch}{1.6}
\begin{tabular}{p{0.44\textwidth} p{0.44\textwidth}}
\hline
\multicolumn{1}{c}{\textbf{Old Assessment Paradigm}} & \multicolumn{1}{c}{\textbf{New Assessment Paradigm}} \\
\hline
Syntax Memorization & Trade-off Analysis \& Architectural Thinking \\
Line-by-Line Authorship & Problem Framing \& Constraint Design \\
Plagiarism Checks for Academic Integrity & Process Transparency \& Oral Defense \\
Isolated Ethics Elective & Cross-Cutting Responsible AI Integration \\
Solo Implementation Projects & AI-Assisted Workflow with Human Oversight \\
\hline
\end{tabular}
\end{table}

\subsection{Tooling: Infrastructure for Orchestration and Verification}
\label{subsec:tool_transform}

The development environment must evolve from code-centric editors to AI-mediating platforms that support orchestration and verification as first-class activities. The era of autocomplete-style assistance (exemplified by early GitHub Copilot) is giving way to full orchestration platforms with built-in verification, provenance tracking, and human-in-the-loop controls~\cite{banh2025copiloting}. This transformation encompasses several interrelated infrastructure requirements.

AI orchestration platforms must provide integrated environments for coordinating multiple agents, managing task delegation, implementing cross-agent verification, and maintaining audit trails of agent interactions and outputs~\cite{mosaico2024}. The rapid maturation of multi-agent frameworks---with standardized communication substrates such as the Model Context Protocol for tool interaction and Agent-to-Agent protocols for peer coordination---is establishing the technical foundation for such platforms~\cite{ren2025agents}. These platforms must treat AI as a ``team member'' with defined roles, not a black box: humans define goals, agents plan, execute, and iterate, and an orchestration layer coordinates perception, memory, action, and collaboration.

Verification infrastructure must deliver rapid-feedback ecosystems featuring ephemeral test environments, advanced static and dynamic analysis capabilities, and continuous semantic validation against formal or semiformal specifications~\cite{dolcetti2025verification}. The dual challenge of using LLMs as verification tools and verifying the code produced by LLMs is now receiving sustained research attention, with emerging approaches integrating LLMs with traditional static analyzers and formal verification frameworks through prompt engineering techniques and hybrid neuro-symbolic methods~\cite{casola2025dualverification}. Semiformal reasoning engines---techniques that allow AI agents to assess code functionality without full execution, combining symbolic analysis with probabilistic reasoning---represent a particularly promising direction, with recent explorations achieving high accuracy rates in specific domains~\cite{li2024verification}. As formal verification tools become increasingly AI-assisted, the historically prohibitive cost of mathematical proof may decline enough to make verified code generation practical for mainstream software~\cite{heidrich2025formal}.

Finally, AI-augmented continuous integration and continuous deployment (CI/CD) pipelines must incorporate AI agents for initial code review, automated remediation suggestions, security scanning, and compliance validation, creating closed-loop quality assurance systems that operate at the speed demanded by abundant code generation. Provenance and governance features---built-in tracking of AI-generated code lineage, decision rationale, and trade-offs---must become standard components of the development toolchain rather than afterthoughts~\cite{banh2025copiloting, safeai2024}.

\subsection{Processes: Verification-First and Human-in-the-Loop Lifecycles}
\label{subsec:process_transform}

Traditional agile and waterfall models must adapt to accommodate AI as a core participant in the development lifecycle rather than as a peripheral tool. As Banh et al.~\cite{banh2025copiloting} emphasize, the successful adoption of generative AI depends not on AI's raw capability but on how well it integrates with established development practices. The process transformations indicated by the evidence converge on a verification-first, human-in-the-loop model that redesigns the entire process rather than merely automating subtasks.

Specification-driven development, in which detailed technical specifications---encompassing shared domain models, use cases, business rules, and constraints---are authored by human engineers upfront and serve as ground truth that AI agents implement, offers a promising approach to reducing hallucination-induced errors and maintaining traceability between intent and output~\cite{roychoudhury2025agentic}. This approach has been characterized as an ``AI Unified Process'' in which the human contribution shifts from implementation to artifact definition, and the engineering value lies in the precision of the specification rather than the volume of the code. Multi-agent verification, in which independent AI models cross-check each other's outputs, can catch domain-specific blind spots and reduce single-model bias, though this approach introduces its own coordination challenges~\cite{mosaico2024}. The security implications are particularly acute: empirical evidence demonstrates that iterative AI code refinement can introduce progressive security degradation, with initially secure code accumulating vulnerabilities through successive AI-mediated modification cycles~\cite{fakih2025security}. This finding underscores the necessity of layered assurance---combining automated tests, formal proofs, and runtime monitors---for any AI-generated code destined for production.

Formal handoff procedures and explicit gates for human oversight must be established at critical junctures---architecture sign-off, security validation, and production deployment---to prevent the ``accountability collapse'' that autonomous pipelines risk introducing~\cite{safeai2024}. Validation pipelines must be redesigned to catch AI-specific failure modes, prioritizing semantic correctness, constraint adherence, and system-level integration over syntactic compilation. The empirical evidence that downstream maintainability depends on process quality rather than model capability~\cite{borg2025maintainability} provides strong justification for investing in process infrastructure over model sophistication. Metrics must evolve from lines of code written to the velocity of verified features delivered and to system-level reliability outcomes.

\subsection{Professional Practice: Roles, Metrics, and Governance}
\label{subsec:practice_transform}

The software engineering profession must evolve toward specialized, oversight-oriented roles with corresponding changes in performance metrics and governance structures. Emerging specializations---such as AI Workflow Engineer, PromptOps Specialist, and AI Quality Guardian---reflect the diversification of engineering work away from general-purpose coding toward orchestration, optimization, and enforcement of technical and ethical guardrails~\cite{alenezi2025ai}. Cruz et al.~\cite{cruz2025redefining} confirm through phenomenological analysis that IT professionals already perceive a shift in their professional identities, moving from traditional coders into strategic collaborators who must navigate intelligent automation, shifting role boundaries, and emerging ethical concerns within Industry~4.0 ecosystems.

The daily work of the software engineer is shifting from line-by-line code authorship to a higher-order activity set comprising task decomposition, output verification, component integration, and exception handling. Junior roles increasingly focus on review and collaboration rather than greenfield implementation, while senior roles emphasize strategy, trade-off analysis, and mentoring AI-augmented teams~\cite{russinovich2026redefining}. Performance metrics must evolve accordingly: success should be measured by decision velocity, AI orchestration efficiency, system reliability, and business outcomes rather than by lines of code written or pull request throughput. Job postings already reflect this shift, increasingly prioritizing experience with AI tools and system-level design over pure implementation skills.

Governance frameworks must establish clear human liability boundaries, maintain audit trails of agent decisions, preserve versioned prompt and specification histories, and require compliance reporting for AI-augmented development processes. These governance mechanisms are not bureaucratic overhead; they are the structural means by which organizations ensure that the accountability principle---the irreducible human component identified across the literature~\cite{zhong2025llmse, alenezi2025ai, safeai2024}---is operationalized in practice. Engineers remain ultimately responsible for outcomes, including security vulnerabilities introduced by AI-generated code, and practices must reflect this through explicit upskilling mandates, interdisciplinary collaboration requirements, and systematic safeguards against automation bias~\cite{cruz2025redefining, banh2025copiloting}.

This redefinition does not diminish software engineering---it elevates it. AI handles the mechanical, freeing humans for the creative, integrative, and accountability-bearing work that truly builds enduring systems. Organizations and professionals who proactively invest in these transformations across education, tooling, processes, and governance will be positioned to thrive; those who cling to the traditional code-authorship paradigm risk both competitive disadvantage and the accumulation of unverified technical debt at unprecedented scale.

\section{Discussion}
\label{sec:discussion}

This section synthesizes the findings by explicitly addressing the four research sub-questions and situating them within the broader systemic transformation outlined in Section~\ref{sec:transformations}. We then examine implications for research and practice, followed by a critical reflection on limitations and threats to validity.

\subsection{Answers to the Research Questions}
\label{subsec:rq_answers}

\paragraph{RQ1: How is the role of the software engineer evolving in the context of agentic AI workflows?}

The evidence consistently indicates a structural redefinition of the software engineer from code producer to \textit{orchestrator of sociotechnical systems}. In agentic environments, implementation is increasingly delegated to AI agents, while human engineers assume responsibility for intent specification, architectural coherence, and outcome accountability. This shift extends beyond a simple redistribution of tasks; it repositions engineering as a discipline centered on \textit{decision-making under uncertainty}.

Engineers now operate at multiple abstraction layers simultaneously: they decompose problems, encode constraints into specifications or prompts, coordinate multi-agent workflows, and adjudicate ambiguous or high-risk outcomes. As Section~\ref{subsec:process_transform} highlights, the locus of value shifts from writing code to ensuring that generated systems are correct, aligned, and sustainable. Crucially, accountability does not diffuse with automation. Instead, it concentrates: engineers remain the final arbiters of system behavior, particularly in edge cases where AI systems fail unpredictably.

\paragraph{RQ2: What verification challenges arise from the abundance of AI-generated code, and what hybrid approaches address them?}

AI-generated code introduces a distinct failure profile characterized by semantic inaccuracies, hidden security vulnerabilities, and misalignment with implicit system constraints. Unlike traditional defects, these failures often evade superficial review due to their syntactic plausibility. At scale, the abundance of generated code transforms verification from a downstream activity into a systemic bottleneck.

The literature converges on \textit{verification as infrastructure} rather than as a discrete phase. Effective responses take the form of layered, hybrid pipelines that integrate LLM-based reasoning with static analysis, dynamic testing, and formal methods. Multi-agent cross-verification, specification-driven validation, and semiformal reasoning engines collectively address different failure modes, but no single technique is sufficient in isolation.

Importantly, the challenge is recursive: LLMs are both generators and verifiers. This dual role necessitates careful trust calibration and governance to prevent correlated failure modes across models. As a result, the engineer's responsibility expands to designing verification ecosystems that are resilient, auditable, and aligned with system-level requirements.

\paragraph{RQ3: What does the empirical evidence reveal about the productivity, quality, and maintainability effects of AI-augmented development?}

Empirical findings reveal a divergence between \textit{local productivity gains} and \textit{system-level outcomes}. While AI assistance can significantly accelerate task completion, these gains are unevenly distributed and highly dependent on developer expertise and task complexity. More importantly, increased speed does not inherently translate into improved quality or maintainability.

The evidence suggests that maintainability and long-term system health are functions of process quality rather than generation capability. Without robust verification pipelines, architectural discipline, and governance mechanisms, rapid code generation risks amplifying technical debt and security exposure. Conversely, when embedded within well-designed processes, AI can enhance consistency, documentation, and test coverage.

This reinforces a central insight: AI shifts the performance frontier but does not eliminate fundamental engineering trade-offs. Productivity gains are real, but they are conditional on the presence of complementary capabilities in verification, integration, and oversight.

\paragraph{RQ4: What transformations in education, tooling, processes, and professional practice are required to operationalize this paradigm shift?}

The transition to AI-augmented engineering requires coordinated transformation across four tightly coupled dimensions:

\begin{itemize}
    \item \textbf{Education:} The emphasis must shift toward systems thinking, verification literacy, and human-AI collaboration. Graduates must be trained to evaluate and govern AI outputs, not merely produce code. Assessment models must prioritize reasoning transparency and defensible decision-making.
    
    \item \textbf{Tooling:} Development environments must evolve into orchestration platforms that integrate agent coordination, verification pipelines, and provenance tracking. Toolchains must treat AI as an explicit participant with observable behavior rather than an opaque assistant.
    
    \item \textbf{Processes:} Lifecycles must become verification-first, with specification-driven development, continuous validation, and explicit human oversight gates. Process design, not model capability, is the primary determinant of system reliability.
    
    \item \textbf{Professional Practice:} Roles, metrics, and governance structures must align with an orchestration-centric model. Performance should be measured by decision quality, system reliability, and effective AI utilization rather than code volume.
\end{itemize}

These transformations are interdependent. Partial adoption risks creating misalignment, where increased generation capacity outpaces an organization's ability to verify, integrate, and govern outputs.

\subsection{Implications for Research}

The findings highlight a need to reorient research from isolated model performance toward \textit{end-to-end system behavior}. First, benchmarking frameworks must evolve to evaluate orchestration and verification pipelines, incorporating metrics such as reliability under iteration, security robustness, and lifecycle maintainability.

Second, the integration of probabilistic LLM reasoning with deterministic formal methods represents a critical research frontier. Achieving scalable, hybrid verification with meaningful guarantees requires new theoretical models that reconcile these fundamentally different paradigms.

Third, there is a pressing need for longitudinal studies of human-AI collaboration in production environments. Key open questions include how expertise develops in AI-mediated workflows, how teams maintain shared understanding, and how organizations prevent skill atrophy while leveraging automation.

Finally, research should address governance and accountability at scale, including mechanisms for auditability, liability attribution, and compliance in AI-mediated development ecosystems.

\subsection{Implications for Practice}

For practitioners, the central implication is clear: \textit{generation capability must not outpace verification capability}. Organizations should prioritize investment in verification infrastructure, specification quality, and governance frameworks before scaling AI-driven development.

Adoption strategies must account for variability in developer expertise and task context. Senior engineers may realize immediate gains, while less experienced developers require structured support, training, and oversight to avoid productivity drag and quality degradation.

Establishing governance early is critical. Provenance tracking, prompt and specification versioning, and audit-ready workflows should be treated as foundational infrastructure. Retrofitting these capabilities after large-scale adoption is both costly and operationally disruptive.

More broadly, organizations must recognize that AI adoption is not a tooling upgrade but a transformation of the engineering system itself, requiring aligned changes in culture, processes, and incentives.

\subsection{Limitations and Threats to Validity}

This study is subject to several limitations. In terms of \textit{construct validity}, key concepts such as agentic workflows and orchestration remain emergent and lack universally accepted definitions, introducing potential ambiguity in interpretation.

Regarding \textit{external validity}, the review is bounded by a specific time window and set of data sources. The rapid evolution of the field means that new developments may quickly supersede some findings. Additionally, much of the empirical evidence is derived from controlled or large-scale industrial contexts, limiting generalizability to smaller or less mature organizations.

For \textit{internal validity}, the inclusion of multivocal sources introduces variability in rigor. While this approach captures emerging insights, it also incorporates perspectives that may not yet be empirically validated.

Finally, with respect to \textit{conclusion validity}, the proposed transformation framework is synthetic and conceptual. Its practical effectiveness remains to be validated through empirical studies, particularly longitudinal and real-world implementations.

Despite these limitations, the consistency of evidence across independent studies and domains provides strong support for the central conclusion: software engineering is undergoing a systemic transition in which human expertise is re-centered on orchestration, verification, and accountability in AI-mediated environments.

\section{Conclusion and Future Work}
\label{sec:conclusion}

The abundance of code generated by large language models and agentic AI systems is driving a fundamental redefinition of software engineering. The evidence reviewed in this paper converges on a clear conclusion: the discipline must shift its center of gravity from manual code production to the orchestration of multi-agent workflows, the rigorous verification of AI-generated outputs, and the structured governance of human-AI collaboration. This paradigm shift does not diminish the importance of human engineers; rather, it elevates their role toward system-level design, semantic verification, and accountable oversight.

To operationalize this shift, we have proposed a framework organized around four core competencies---intent articulation, systematic verification, multi-agent orchestration, and human judgment---and have detailed the concrete transformations required across education, tooling, processes, and professional governance. Education must pivot from syntax mastery to systems curation, emphasizing verification literacy, architectural thinking, and responsible AI practice. Development tools must evolve into orchestration platforms with embedded verification infrastructure. Lifecycle processes must adopt specification-driven, human-in-the-loop models that treat verification as a first-class activity. And professional practice must redefine success around system outcomes, collaborative fluency, and accountable governance.

Several avenues for future work emerge from this analysis. The most urgent priorities include the development of open standards for prompt provenance and agent capability declaration; the design and empirical evaluation of verification-first curriculum interventions; controlled experiments comparing orchestration-centered and authorship-centered development workflows in industrial settings; and the formal integration of LLM reasoning with symbolic verification methods. Longitudinal field studies tracking the career trajectories and skill evolution of engineers working in AI-augmented environments will be essential for validating the workforce transformation predictions advanced in this and related work.

By embracing these transformations with both ambition and rigor, the software engineering community can harness the abundance of AI-generated code responsibly---ensuring that the discipline remains human-centered, trustworthy, and capable of delivering sustainable societal value.

\bibliographystyle{ieeetr}
\bibliography{name}

\end{document}